\documentclass{article}
\usepackage{fullpage}
\usepackage[utf8]{inputenc}
\usepackage{color}

\usepackage{natbib}
\title{Practical Considerations for Differential Privacy}
\date{}
\author{Kareem Amin\hspace{0.3in}
\and Alex Kulesza
\\ \{kamin, kulesza, sergeiv\}@google.com
\\Google Research, NY
\and Sergei Vassilvitskii
}

\begin{document}
\maketitle

\begin{abstract}
    Differential privacy is the gold standard for statistical data release. Used by governments, companies, and  academics, its mathematically rigorous guarantees and worst-case assumptions on the strength and knowledge of attackers make it a robust and compelling framework for reasoning about privacy.  However, even with landmark successes, differential privacy has not achieved widespread adoption in everyday data use and data protection. In this work we examine some of the practical obstacles that stand in the way.
\end{abstract}

\vspace{0.25in}

Differential privacy has seen significant practical success in recent years, with high profile adoptions by the US Census, Covid reports, and industry in general~\citep{census,covid,plume,opacus}. It has also been extensively studied from a theoretical standpoint, and the insight that the privacy of a system can be measured by the effect of a single individual on its output has been widely influential. 

However, the core promise of differential privacy---that it can ``make confidential data widely available for accurate data analysis,
without resorting to data clean rooms, data usage agreements, data protection plans, or restricted views''~\citep{dworkandroth}---remains unfulfilled in practice. Of course, minor frictions and mismatches are inevitable when any formalism meets the real world. Previous work has reported on the difficulties in selecting epsilon, communicating the guarantees to various stakeholders, designing high-utility algorithms for realistic problems, and launching deployments in machine learning settings~\citep{blancojusticia, cliftontassa, domingo2021limits, drechsler2023differential, garfinkel2018issues, reiter2019differential, cummings2023advancing, cummings2023centering}. Our focus is not to re-hash the difficulties that have already been well-documented in these works, but rather, to offer an explanation for why these challenges are so pervasive in practical deployments. We argue that, due to its intentional focus on worst-case outcomes, differential privacy faces a deeper, more fundamental challenge: applying it successfully in practice paradoxically requires answers to some of the same difficult and messy privacy questions that differential privacy was meant to obviate.


We present three examples of this phenomenon, each arising at the interface between the mathematical formalism of differential privacy, which demands strict control over sensitivities, accumulated privacy loss, and neighborhood relations, and the realities to which it is frequently applied.

\section*{Bounding Sensitivity}

Because differential privacy is concerned with worst-case information leakage, most (though not all) mechanisms in the literature require that data comes with known bounds. For instance, mechanisms for computing the mean require knowing the range of values~\citep{dworkandroth}, and mechanisms for clustering often assume that all points lie in a unit ball~\citep{Clustering}. Perhaps even more common is the (implicit) assumption that each individual contributes only a single record to the dataset. 

In practice, such assumptions must be enforced; real data is messy and frequently unbounded. Outside of highly controlled datasets such as the US Census, individuals can often contribute many records, each of which may take on arbitrary values, including genuine outliers as well as error values. Cleaning and bounding the data is therefore usually one of the first problems encountered by a practitioner of differential privacy. Unfortunately, designing a pre-processing routine to enforce these requirements can be fraught, not only because it has outsize effects on data quality and computational implications, but also because it demands difficult judgments that fall outside the bounds of differential privacy.




For example, suppose we need to impose a limit on the magnitude of the values in a dataset. As has been well documented, long tails in real data distributions are the rule, not the exception, so the maximum and mean values in the data are usually far larger---possibly orders of magnitude larger---than the median. This creates a fundamental tension between bias and variance when choosing the cap. We can impose stringent bounds, which will keep downstream noise levels due to differential privacy low but, by cutting off the tail of the distribution (where it may be most anomalous or interesting), introduce significant bias. Alternatively, we can impose loose bounds, preserving more of the data but forcing the noise to be potentially orders of magnitude larger than what is required to protect a ``typical'' individual. This high-variance solution ultimately generates unreliable signals. While there is some theoretical work on the effects of contribution bounding, as well as methods for implementing them~\citep{biasvariance, followup, msft}, this trade-off remains a problem in practice. The question of {\em where} the bounds come from---whether they are somehow discovered by the analyst, or calculated in a private way from the data---is often elided.  

And yet, this is a critical choice if we aim to control overall privacy leakage.
 If a contribution cap is derived from the dataset in a manner that is not differentially private, whether directly or through the exploration and experience of an analyst, then the formal bounds of the downstream mechanism are negated.
 It may be tempting to simply exclude the choice of cap from privacy loss assessment on the basis that it is ``reasonable'' or unlikely to create a meaningful attack surface---both may be true---but such judgments are inherently at odds with the philosophy of differential privacy, which aims to provide ``protection against arbitrary risks''~\citep{dworkandroth} and quantify privacy loss without human judgment. 

The need to perform contribution bounding creates other privacy trade-offs as well. In order to bound data correctly, it is necessary to know which records were contributed by the same real-world individual. To achieve this, we must associate stable identifiers (such as email addresses, user IDs, cookies, device IDs, or pseudonymous derivatives) with records in the database. This creates a tension between differential privacy and data minimization: for correct differential privacy, the curator must maintain an especially rich store of personal information, whereas otherwise such identifiers could be stripped. In the extreme, central DP might require identifiers to be held alongside an unaggregated dataset for its entire lifetime.

Going further, the mapping from stable data identifiers to real-world individuals---the entities we ultimately want to protect---may not actually be one-to-one; a web cookie is not a human being. This distinction might seem unimportant, but individuals more concerned about their privacy are often precisely those who cycle their data identifiers frequently. For instance, privacy-sensitive individuals might be more inclined to start a new private browser session each time they visit an embarrassing website. An operationalization of differential privacy that treats cookies as equivalent to real-world individuals can not only fail to provide the desired level of protection in general but also, ironically, give the \emph{least} protection to those who place a special importance on their own privacy.

Once again, the need for contribution bounding directly introduces questions---about whether data minimization outweighs differential privacy, or how to relate real-world individuals to identifiers---that can only be answered by wrestling with difficult issues outside the walled garden of differential privacy.


\section*{Privacy Loss Accounting}

In an idealized world, DP algorithms could be deployed everywhere, allowing for perfect reasoning about the privacy loss of a system (measured in terms of $\epsilon$ or some other accounting parameter). In practice, there are significant challenges to realizing this vision even approximately.

First, privacy loss accounting is not conducive to exploration. The DP literature generally imagines a world where the data analyst arrives with a fixed goal. For example, if they want to compute a mean, they can deploy algorithms designed to extract the most accurate estimate given a known privacy budget. Contrast this, however, with a real data analyst's typical workflow, which may involve making dozens of exploratory queries to understand the dataset, design appropriate filtering policies, and debug and refine query logic before they have even settled on the question they would like to ask. While this topic has received some limited attention~\citep{dwork2018fienberg} in the literature, expanding the scope of practical solutions to this problem is critical to the success of DP. 

Given existing difficulties in enforcing hard privacy budgets, a natural alternative could be to replace budgeting with monitoring. That is, instead of restricting the usage of a DP system when budgets have been exhausted, one could simply \emph{track} the accumulated privacy loss of system, hoping to drive the rate of loss down over time with algorithmic improvements, incentives, and so on. Proponents of this viewpoint must contend, however, with the fact that privacy loss grows very quickly in real systems. In the example above, a single analyst using the system in their normal manner might easily generate a privacy loss of $\epsilon = 50$ or more in a single session, even using sophisticated composition techniques. At this level of privacy loss, let alone the accumulated loss of multiple sessions and/or multiple analysts over time, the mathematical guarantee of DP is all but meaningless.


One might hope that, despite the weak formal bound, DP can still prevent real dangers such as dataset reconstruction attacks or memorization attacks. But if we are to rely on resistance to empirical attacks to justify our privacy systems, then we must also question what value the mathematical abstraction of DP provided in the first place. While some patterns of usage might be acceptable at $\epsilon=50$, the obvious existence of \emph{unacceptable} mechanisms with the same guarantee suggests that accumulated differential privacy loss may simply not be well-aligned with real-world goals. Today we don't have a good vocabulary to distinguish the two scenarios. 

Finally, we must accept that, in any sufficiently large system, it is almost a certainty that a deterministic function of user data will be computed and published somewhere. It may be as a byproduct of the larger system (for instance, a resource management system sharing total size of the dataset on disk, rounded to the nearest terabyte) or from other analyses of the same data using a different privacy model (see ~\cite[Section 4.1.3]{reiter2019differential} for an example).  If we are to track privacy loss, we must decide how to handle such deterministic releases. On the one hand, we can be strict and include them in our privacy accounting, uselessly concluding that the system has infinite privacy loss. Or we can exclude them, which, if we are to be rigorous, requires establishing some principle by which certain releases are subject to loss accounting and others are not. This, of course, begs the very question that DP was intended to answer in the first place: how can we determine if a data release is privacy-safe?

\section*{Neighborhood Definitions}

Faced with the challenges of extracting utility under differential privacy, researchers have taken to adapting or relaxing the definition of differential privacy to settings where mathematical rigor is appreciated, but the requirements of traditional differential privacy are not a good fit. While these adaptations are sometimes driven by necessity, it is important to examine both the formal and real-world privacy protections they provide.

One part of the definition that is often relaxed is the neighborhood relation. Standard differential privacy considers datasets $D$ and $D'$ neighbors if they differ by the contribution of a single individual, colloquially referred to as ``user-level'' differential privacy.  However, there are many alternative ways to define the concept of neighboring databases (see \citet{Desfontaines} for examples). For instance, in machine learning applications it is common to use so-called ``record-level'' DP, defining $D$ and $D'$ to be neighboring if they differ by one example, even if a user can contribute multiple examples. 

Relaxing the neighboring relation is sometimes used as a flexible technique for addressing the challenges described in the previous sections. For periodic releases, we can sidestep the problem of ever-increasing privacy loss by defining a neighboring dataset as one that differs by one user during some limited time window (such as a single day). We can similarly sidestep the disconnect between identifiers and real-world individuals by defining neighboring databases in terms of identifiers and not human beings. However, in each of these cases the fundamental problem has not really been resolved. The syntactic guarantee for more granular privacy units may be called ``differential privacy'' but does not, by itself, materially change the consequences in the real world. 


This type of mismatch between formal guarantees and the social implications of a release manifests beyond just changing the granularity of the neighboring relation. For example, in label DP~\citep{LabelDP} each record is taken to be a  multi-dimensional vector, and all but the last coordinate are assumed to be public knowledge. These public dimensions are ``features'', and the last coordinate is the ``label'', which is private to the user. The goal of label DP is to allow computations on the entire database while protecting only the \emph{label} of each user in the differential privacy sense.

It is easy to describe formally the protections that label DP provides. However, the privacy implications for an individual may be quite different if there are strong correlations between the features and the label that were previously unknown. While statistical inference is not a privacy violation under the standard of differential privacy (for example, learning that ``smoking causes cancer''  without learning the smoking or cancer status of any individual is a desirable goal), the amount of plausible deniability retained by an individual under label DP depends on the strength of these correlations, and \emph{not} (in general) on the specific value of $\epsilon$ used to extract the hypothesis from the data, as noted by \citet{BusaFekete2023} and \citet{LIA}. 

In any case, the assessment of whether label DP provides sufficient privacy requires consideration of empirical correlations that sit well outside the scope of traditional differential privacy. Similarly, relaxing the neighboring relation might well be consistent with privacy expectations in a particular system, but the need to evaluate this claim at all cuts against the promise of differential privacy, which is to remove subjectivity in comparing privacy risks, and to quantify privacy loss under a definition that does not need to be reexamined with each new deployment.

\section*{Conclusion}

Differential privacy provides a robust and formal definition for reasoning about the fundamentally social concept of privacy. We might hope to elevate conversations about privacy using that definition, moving from a world of soft trade-offs, heuristics, human considerations, and philosophical disagreements to one of clarity, reason, theorems, algorithms, and rigor. In practice, this dream is far from a reality. Although there has been progress---for instance, work on private parameter tuning~\citep{liu2019private}, development of new algorithms that avoid precise knowledge of data bounds~\citep{biswas2020coinpress,brown2024insufficient}, and the aforementioned work on the Fienberg problem ~\citep{dwork2018fienberg}---practitioners still regularly wrestle with difficult privacy questions outside the scope of differential privacy. And, as we have argued, differential privacy itself often forces them to do so.



Today, practitioners resolve these questions using a combination of intuition, experimentation, convention, and regulatory and/or legal guidance. However, none of these is philosophically satisfying in the way that differential privacy can be. To move forward, we need to bring rigor to the informal reasoning embedded in real-world deployments, especially in cases where ``weak'' differential privacy (using non-private bounds, large budgets, or relaxed neighborhood definitions) is believed to provide protections that have yet to be carefully articulated and recognized.

\section*{Acknowledgements}
We thank Adam Smith, Jalaj Upadhyay, Abhradeep  Thakurta, and Thomas Steinke for fruitful discussions and comments on earlier drafts of this work. 


\end{document}